Directed evolution effectively selects for DNA based physical reservoir computing networks capable of multiple tasks

Tanmay Pandey[1,2], Petro Feketa[3,4], Jan Steinkühler[1,4*]

[1] Bio-Inspired Computation, Institute of Electrical and Information Engineering, Kiel University, Kiel 24143, Germany

[2] Department of Biological Sciences, Indian Institute of Science Education and Research, Mohali, Knowledge City, SAS Nagar, Manauli PO 140306, India

[3] Chair of Automation and Control, Institute of Electrical and Information Engineering, Kiel University, Kiel 24143, Germany

[4] Kiel Nano, Surface and Interface Science KiNSIS, Kiel University, Kiel, Germany

* Corresponding author jst@tf.uni-kiel.de

## Abstract

DNA and other biopolymers are being investigated as new computing substrates and alternative to silicon-based digital computers. However, the established top-down design of biomolecular interaction networks remains challenging and does not fully exploit biomolecular self-assembly capabilities. Outside the field of computation, directed evolution has been used as a tool for goal directed optimization of DNA sequences. Here, we propose integrating directed evolution with DNA-based reservoir computing to enable in-material optimization and adaptation. Simulations of colloidal bead networks connected via DNA strands demonstrate a physical reservoir capable of non-linear time-series prediction tasks, including Volterra series and Mackey–Glass chaotic dynamics. Reservoir computing performance, quantified by normalized mean squared error (NMSE), strongly depends on network topology, suggesting task-specific optimal network configurations. Implementing genetic algorithms to evolve DNA-encoded network connectivity effectively identified well-performing reservoir networks. Directed evolution improved reservoir performance across multiple tasks, outperforming random network selection. Remarkably, sequential training on distinct tasks resulted in reservoir populations maintaining performance on prior tasks. Our findings indicate that DNA-bead networks offer sufficient complexity for reservoir computing, and that directed evolution robustly optimizes performance.

**Introduction**

To design more energy-efficient and resilient alternatives to digital, silicon-based computation, new types of "in-material" computation have been proposed.(1,2) A promising direction is neuromorphic sensing: by importing event-driven transduction, adaptive gain control, and predictive coding from biology, we can perform low-latency, low-power computation at the sensor itself.(3–5) As a substrate for neuromorphic sensing, biological macromolecules are particularly interesting because of their biocompatibility, ability to self-assemble, cheap production and sustainable sourcing. For example, DNA strand based computing for classification,(6,7) Hopfield-like associative memory,(8) physical learning in soft and active matter,(1) and reservoir computing approaches(9–14) have been studied. A common problem to all these approaches is the design of a network structure and biomolecular interactions that optimally solve a given task. Additionally, systems devised so far are mostly static according to the original top-down design.  In neuroevolutionary and related fields like genetic programming, evolutionary principles such as selection and growth are considered.(15) Task-performance has been also used to select growing reservoir computing networks that yield better performance than random Erdős–Rényi graphs(16) and networks selected for task-performances show signatures of critical dynamics. (17) Some earlier approaches to evolution of reservoirs were reviewed.(18) However, these approaches are limited to computing architectures evaluated on digital computers. At the same time, DNA sequences, biomolecular binding, and other biochemical reactions have been shown to be efficiently optimized by the process of directed evolution (19) - that is, the selection of physical systems based on task performance. In this work, we propose a combination of directed evolution and reservoir computing using DNA as a substrate to perform network optimization and adaptation in-material. In physical reservoir computing a material serves as a high-dimensional non-linear projection to perform classification, time series prediction, or other computational tasks.(20–22) Reservoir computing is often applied with physical computing substrates because it only requires adjusting weights of the output layer and the material can remain as prepared, often in a random state determined by its preparation history. Here we focus less on the training and learning methodology and more on evolution of the substrate structure. Comparing the different approaches, there is currently no concept of a reservoir computer that is assembled bottom-up, operates close to real time, and is capable of in-material optimization and evolution (**SI Table 1**).

As a possible implementation, we study the simulation of a network of colloidal beads connected by DNA strands. Because in the proposed system the connections between the beads are coded by the DNA sequence, the same DNA material that provides the substrate for computation can undergo *in-vitro* directed evolution. The goal of this study was to understand the feasibility of the here proposed system in a controlled simulation environment, characterize its capabilities, and deduce quantitative information for a subsequent realization. This work is structured around these ideas: first, we describe the physics-based simulation of the DNA-bead system and investigate its use as a reservoir for time-series prediction. We then propose a coding strategy for the DNA-bead network structure that is compatible with directed evolution. Finally, we simulate the adaptation of the network via directed evolution and evaluate the system under a sequence of tasks to examine its history-dependent performance.

**Results and Discussion**

*A physical reservoir based on DNA-bead networks*

Based on earlier work on reservoir computing with spring networks by Hauser et al., we considered a model of the proposed system based on integration of Newton's equations of motion.(23) In contrast to the original work of Hauser et al., which considered a macroscopic spring–mass system, we performed simulations of mesoscopic colloidal systems using Brownian dynamics. (24) Specifically we modeled colloidal beads functionalized with single-stranded DNA with "sticky-ends" for sequence-specific hybridization and binding of beads. We chose a DNA-bead based material as model because it can be synthesized and manipulated with established methods and is well described by Brownian dynamics.(25–29) In the simulation, the bound beads interacted via the worm-like-chain model of DNA polymer, acting as a non-linear spring between the beads (**Figure 1a**, see Methods for further details). Two beads were fixed in their positions to stabilize the structure in the two-dimensional plane. To add some heterogeneity to the simulation, relaxed DNA spring lengths were drawn at random between 1 µm and 200 µm. Initially, we considered random connectivity between the beads, resulting in a disordered network. As input to the system, we changed the position of a single "input" bead. This movement propagated through the network of DNA-beads by overdamped dynamics corresponding to the low Reynolds number regime. For simplicity, our simulation did not consider thermal fluctuations or other noise sources. We studied a network of $N$ = 8 beads that forms a binary undirected graph with a maximum of 28 different edges and thus a total of $2^{28}$ different topologies.

Because of the non-linear response and fading memory of the DNA springs, the network might be considered for reservoir computing. For this, the weighted linear combination of observables was adjusted to minimize the error between input $X(t)$ and desired output signal $\widehat{Y}(t)$. As reservoir observables we used the spring lengths (see Methods for details). In an experiment, the position of the input bead would be modulated by optical tweezers, and the position of the beads tracked by fluorescent microscopy(12,30). Based on earlier work, we considered two nonlinear memory tasks based on Volterra operator series (Tasks 1 and 2), and one chaotic prediction task (Task 3).(23,31) Specifically, Tasks 1 and 2 were a second-order Volterra operator with a Gaussian kernel $h_2(\tau_1, \tau_2)$:

$$(1)\ \widehat{Y}(t) = \int_0^\infty \int_0^\infty h_2(\tau_1, \tau_2) U(t-\tau_1) U(t-\tau_2) d\tau_1\, d\tau_2$$

Where $U(t)$ is a multiplication of three sinewaves with fixed frequencies as an example signal and varying kernels $h_2$ for the two tasks (**SI Table 2)** for parameters and kernel plot). For both Task 1 and Task 2, the input signal chosen was $X(t) = U(\text{t})$ and target function for determining the weights of the output layer was $\widehat{Y}(t)$. Task 3 was the chaotic Mackey–Glass equation with $\tau = 17$:

$$(2)\ \frac{du}{dt} = a \frac{u(t-\tau)}{1 + u(t-\tau)^c} - bu(t)$$

The input chosen was $X(t) = u(t)$ and the target signal was $\widehat{Y}(t) = u(t+1)$ (**SI Table 3**). Sample trajectories are shown in SI **Figure S1** (Supplementary Information). For a given reservoir structure, the performance was quantified for each task by the normalized mean squared error (NMSE) between output and target signal (**Figure 1b**), even the relatively small reservoir networks considered here approximated the signal well.

Because the here performed simulations should demonstrate feasibility of a physical implementation, we checked for the sensitivity of the reservoir realization. First we considered the effect of the initial random distribution of spring parameters, e.g., due to variations in the bead placement or DNA synthesis. For a fixed topology, the NMSE only varied below 1% with random initialization of the bead positions (**SI Figure 2**). Further we studied the influence of Brownian noise at room temperature and found that this effect degraded performance by less than 20 % for all three tasks (**SI Table 4**). For a physical implementation the possible topologies should be differentiated between planar and non-planar networks. Non-planar networks would require two DNA strands to cross in space (like one connection shown in Figure 1a) when the beads are confined to a 2D surface. So far only non-crossing 2D DNA-bead networks have been demonstrated experimentally, even if the separation of length scales between beads and DNA should make crossing connections possible. We have compared the performance of planar and non-planar networks and found similar performance, indicating that limitation to planar topologies would not prohibit reservoir computing with DNA-bead networks in principle (**SI Figure 3**). This analysis indicates that reservoir computing with the DNA-bead networks suggested here is stable against variations in its in-material realization and therefore a feasible reservoir substrate.

*Effects of DNA-bead network connectivity on reservoir performance*

We found that (microscopic) network connectivity had a large effect on NMSE, as seen by the distributions of NMSE values for 1000 random Erdős–Rényi graphs with probability $p = 0.5$ to form an edge (**Figure 1c**). In addition, the three tasks appeared to differ in difficulty and sensitivity to the network structure, as indicated by the varying average NMSE values and their relative distribution within one task. This indicates that each task has a preferred reservoir network structure. Similarly, the memory and prediction tasks seem to have a varying preference for the magnitude of dynamic node displacements over time, seen by comparing the worst, best, and average performing networks (**SI Figure 4).** These preferences limit the performance of networks for changing tasks: The best-performing network for task 1 (System A, Table 1) showed only average performance for tasks 2 and 3. Similarly, an all-to-all connected network performed did not perform ideally on task 1 but very well on tasks 2 and 3 (Table 1). To understand which fraction of networks performed well on multiple tasks, we correlated the NMSE for two tasks (Figure 2d). This showed that for the majority of randomly drawn networks, performance on one task alone was a poor predictor for performance on other tasks. Only $0.3\%$ of networks were in the top $10\%$ for all three tasks (**SI Table 5**). These results show that small DNA-bead networks are in principle suitable for use in reservoir computing for both time series prediction and Volterra operator tasks but would benefit from optimization of the network structure before the learning of the output layer weights.

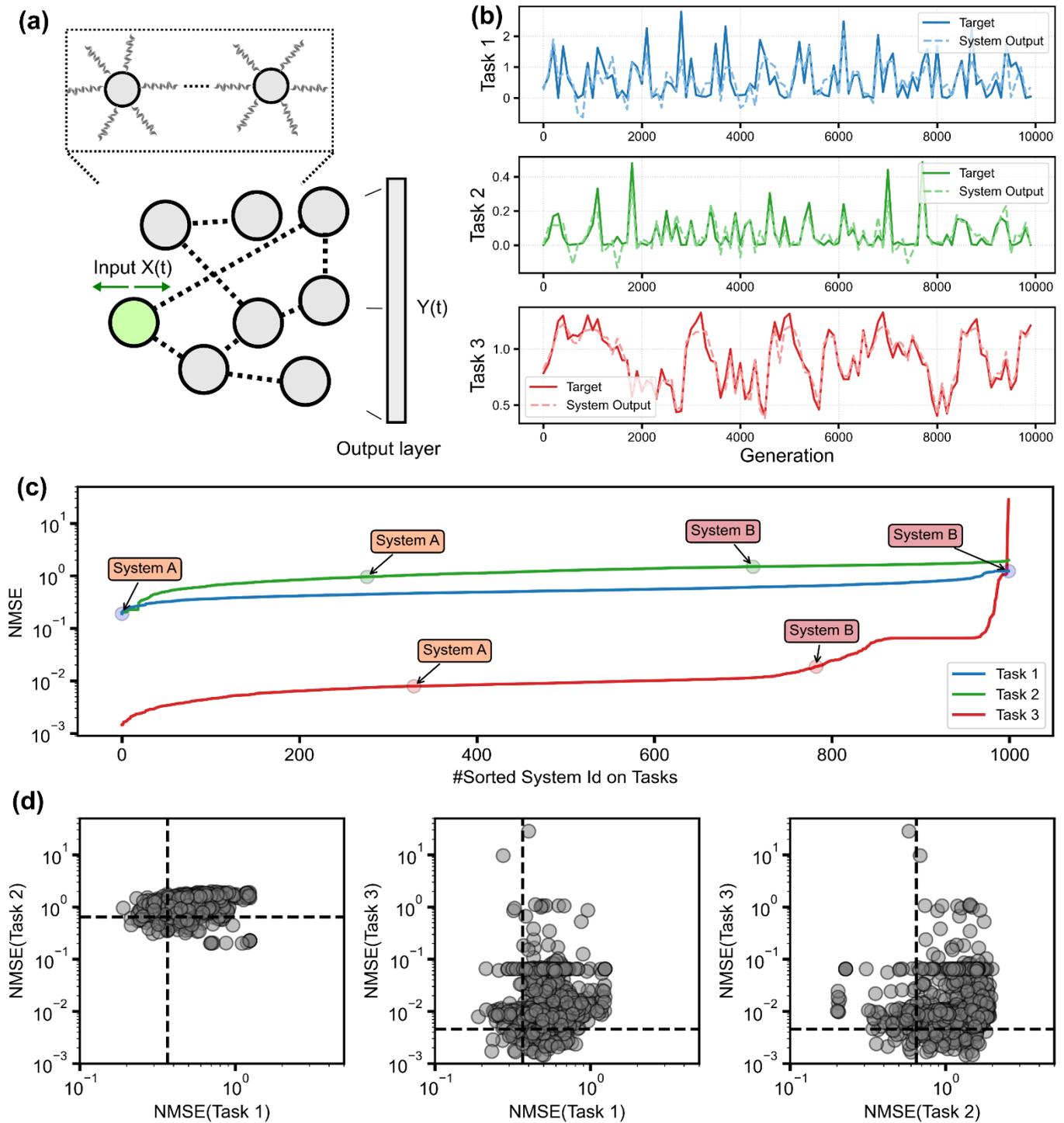

**Figure 1.** a) Sketch of the simulated bead-DNA spring network based on sequence-specific binding of two colloidal beads by DNA with sticky ends grafted on the beads (dashed box inset). The non-linear DNA springs (thick dashed lines) transform the input (position of green bead) into a higher-dimensional representation so that weighted linear combination is the output $Y(t)$. b) Examples of randomly generated network output $Y(t)$ (dashed lines) after training for three different tasks (see main text). c) Normalized mean squared error (NMSE) for the three tasks for 1000 random networks sorted by their NMSE (from small to large). Indicators show selected networks discussed in the main text. d) Correlation of individual networks NMSE (each datapoint) of randomly generated networks shows that NMSE on a single task is a bad predictor for NMSE on a different task. Dashed lines show top 10% NMSE defined by distribution of NMSE on each task.

| System | NMSE on task 1 | NMSE on task 2 | NMSE on task 3 |
|---|---|---|---|
| **System A** | 0.1898593 | 0.9597274 | 0.0079115 |
| **System B** | 1.2375369 | 1.4995228 | 0.0186702 |
| **System C** | 0.3126565 | 0.394345 | 0.0013822 |

**Table 1.** The performance (normalized mean-squared-error) of system A (the best performing system for task 1), system B (the worst performing system for task 1), and System C (an all-connected network) across all different tasks used.

*Directed evolution to select well performing networks*

Next, we considered if directed evolution could be used to select optimal networks. By construction, the connectivity of the studied system can be coded in a DNA string $s$, the "genome". In the proposed coding, each bead is barcoded by a two-letter code $b_i$, e.g. TA or AG. A four-letter substring $s_c$, e.g., TAAG, then codes for the presence of a connection between beads $b_{i,j}$ (**Figure 2a**). Importantly, in this coding, the non-coding substring TA**C**G requires only one mutation operation to transform it into a connection coding substring. Similarly, the same connection might be coded by multiple copies of the same subsequence $s_c$, giving it robustness against deletion. When these "genotypes" are mapped onto reservoir "phenotypes" $\Re(s)$ we found that small differences in genome coding string could yield orders-of-magnitude differences in task performance (**Figure 2b** and **c, Table 2**), indicating that the suggested coding is effective to sample a wide range of reservoir networks.

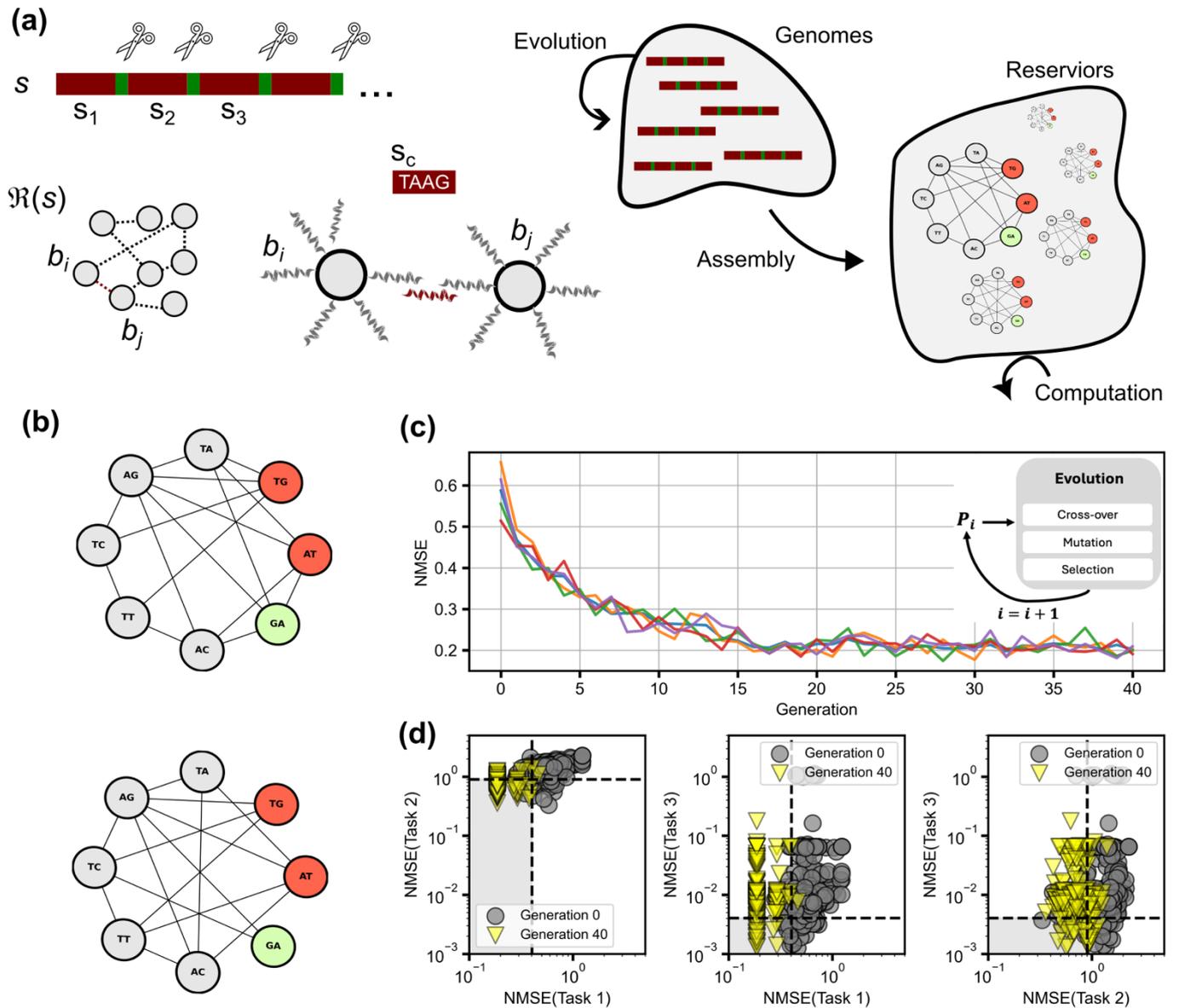

**Figure 2 – (a)** Sketch of network coding string *s* that via (enzymatic) cutting into substrings codes for individual connections. As an example, the cross-linking adapter TAAG between the complementary strands on beads $b_{i,j}$ is shown. Each string *s* codes for a reservoir. **(b)** Networks of the best (top) and worst (bottom) performing individuals on Task 3 (see Table 3 for coding DNA strings) **(c)** Mean NMSE across generations for five independently evolved populations of 300 individuals on Task 1. Inset shows evolution of population $P_i$ of individual

reservoirs $\mathfrak{R}_s$. **(d)** NMSE tasks correlation plots for one such population. Datapoints show individual networks from initially random genome (generation 0) and after task-performance selection (generation 40).

|  | **Best** | **Worst** |
|---|---|---|
| **Coding Sequence** | ATTATCTTGAAGAGTCTTTGACTTAGATTGAGAGTAGAACTTAAGATATATGTGTCATACACAGGAAT | ATTATCTTGAAGAGTCTTTGACTTAGATTGAGAGTATAACTTATGATCTAGGTGTCATACACAGGGAT |
| **NMSE** | 0.0001936 | 0.6877731 |

**Table 2.** The NMSE for the best-performing individual and the worst-performing individual from Figure 2 (a,b).

The coding of the strings considered is not purely symbolic. In an in-material realization of this system the strings $s_1, s_2, s_3, \ldots$ would be coded by on a single-stranded DNA (ssDNA) $s$ separated by enzymatic scission sites (indicated by scissors in **Figure 2a**). Addition of e.g. Cas14 enzyme with corresponding guide RNA would generate individual ssDNA sub-strings(32). Individual substrings would bind the complementary sticky ends of the DNA functionalized beads in a sequence-specific manner by base-pairing. Such DNA based linkers between beads have been studied in theory and experiment before(33–35). A wide range of enzymatic and chemical methods for mutation and recombination of DNA strands exists, which vary in their mutation and recombination rates and purity.(19) Thus the DNA that encodes network structure is in principle feasible to undergo in-material evolution.

In this simulation study, we do not consider the details of the molecular implementation of the directed evolution procedure but simulate the mutation and recombination using a genetic algorithm (GA) that operates on genome strings. Each individual network configuration was coded by a 300-long string, and we considered a population $P_i$ of 100 individual strings. An individual physical reservoir simulation $\mathfrak{R}_s$ was then realized with the prescribed connectivity.

For a time-varying input signal, the linear readout was adjusted to approximate the target signal as before and the inverse NMSE fitness $f(\Re) = NMSE^{-1}$ was evaluated for each individual genome. By cross-over, mutation and selection, a new population $P_{i+1}$ was generated (see Methods for details).

We performed these operations for a total of 40 generations on task 1, which reduced the average population NMSE as expected (**Figure 2c**). Notably, the number of generations needed for convergence is rather small, realistic for *in-vitro* directed evolution. We varied the probabilities for crossover and selection and found that convergence did not depend strongly on the exact values of these parameters (**SI Figure 5a,b**), further showing that a physical realization of the proposed system is feasible. Next, we considered the correlative performance of networks selected on task 1 for tasks 2 and 3 (**Figure 2d**). We found that, while performance on task 1 improved, performance on tasks 2 and 3 did not improve proportionally. This meant that the population distribution was shifted in the direction of the x-axis only, a type of premature convergence of the evolutionary selection.(36) These results show that selection based on task performance is an effective (compared to random selection) method to optimize network structure for a DNA-bead-based reservoir towards a specific task. However, excessive selection on one task might lead to reservoir networks that do not improve multi-task performance compared to random selection.

*Sequences of tasks shape evolutionary trajectory*

We hypothesized that selection of multiple tasks might lead to a more diverse population that preserves a type of memory of the previous tasks in the genome. To this end, we considered sequences of tasks that were the basis for selection of reservoir computers $\Re_s$ (**Figure 3a**). Importantly we selected networks only for 10 generations, avoiding premature fixation of the population on one task. We made two main interesting observations: Firstly, a population initially trained on a task can retain its fitness even if subsequently selected for another task. For example, a population initially trained on task 1 remains fit for task 1 even after selection on task 2 or 3 (**Figure 3a,b**). Secondly, for some tasks the temporal sequence is interchangeable, for others it is not. Selection on task 2 produces a population that is fit on task 1 (**Figure 3c**), however selection on task 2 does not improve average performance on task 3 (**Figure 3d**). Importantly, effects were more pronounced along individual trajectories; for example, the green trace in **Figure 3a** shows low NMSE for both tasks 1 and 2. This suggested that some individual selected networks perform much better than average. This effect should become more pronounced in the individual network correlation plots (**Figure 3e**). Indeed, and in contrast to random networks and selection on single tasks (**Figure 1d, 2d**), the NMSE population correlation plots were shifted towards the third quadrant of the plot, indicating a large population of networks that perform well on multiple tasks. This effect can also be seen in the fraction of top 10% networks overlapping between the three tasks (**Figure 3f**). All four task sequences provided a much larger fraction of well-performing reservoir networks than random selection for all tasks correlations (**I, II, III** in **Figure 3f**). This is also true for the challenging intersection of top 10% for all three tasks (**IV** in **Figure 3f**). This shows that selection on tasks (1,2) can even select network that are fit for task 3. Directed evolution with short task sequences selects DNA-bead reservoirs that are adaptable to new tasks without losing the ability to perform well on previous tasks seen during their evolutionary history.

*Network entropy converges with evolutionary selection*

Finally, we considered if selection on task performance leads to a measurable change in macroscopic network properties in a population of reservoirs. Interestingly, the top-performing networks were far from being fully connected, with an average number of four edges per node in the eight-node networks (**SI Figure 6**). Even more clearly, the network entropy calculated from the node degree of well performing networks was found to be closely constrained to moderate values (0.4-0.6) (blue in **SI Figure 7**). This showed that there was a certain optimal level for the interconnection topology entropy. Such constrained entropy values were also observed in the optimized reservoirs of spin-torque nano oscillators,(17) and are well aligned with information theory of complex networks.(37) However, network entropy is not a sufficient predictor of good multi-task performance as a fraction of poorly performing networks also exhibit moderate entropy values (0.4-0.6) (red in **SI Figure 7**). Therefore, optimization by evolution would still be desirable for an in-material realization. We can conclude that such optimization should start from an initially very sparse population to reach many well-performing networks with entropy values around 0.5. This is because sparse networks (i.e., networks with a low number of links) can easily undergo evolutionary changes if the external signal propagates poorly through them, as they quickly gain additional links via evolution. In contrast, all-to-all networks (or networks with a very high number of links) are rather persistent under evolutionary changes if the signal propagates well through the network; such networks do not lose links so easily during evolutionary selection. The insights from the network analysis can therefore guide the design of an in-material realization.

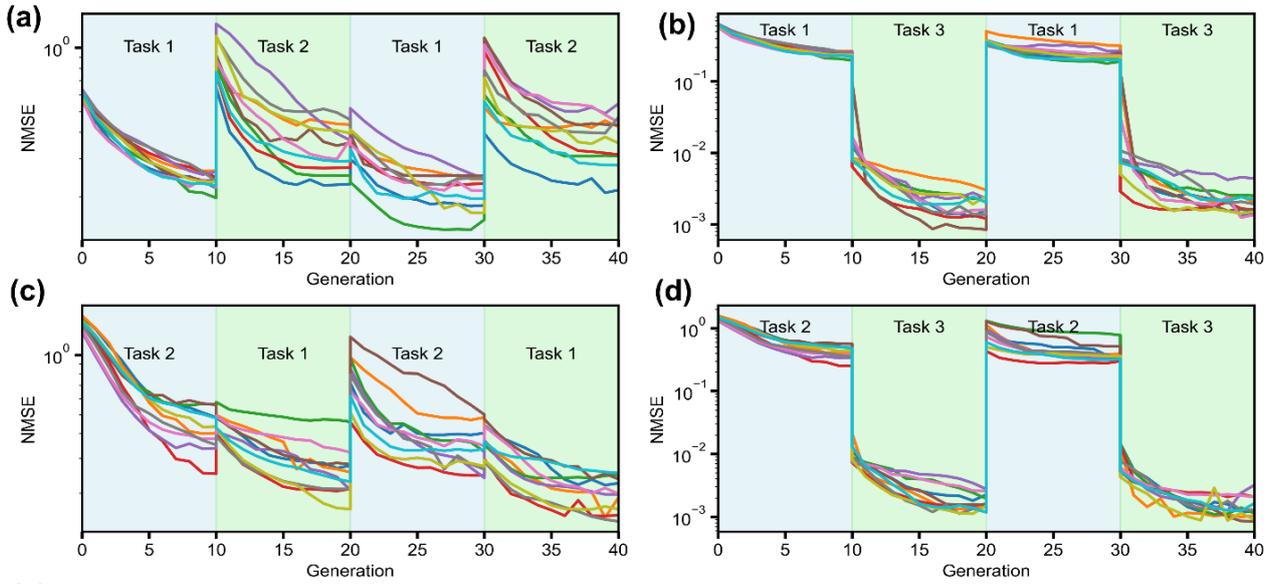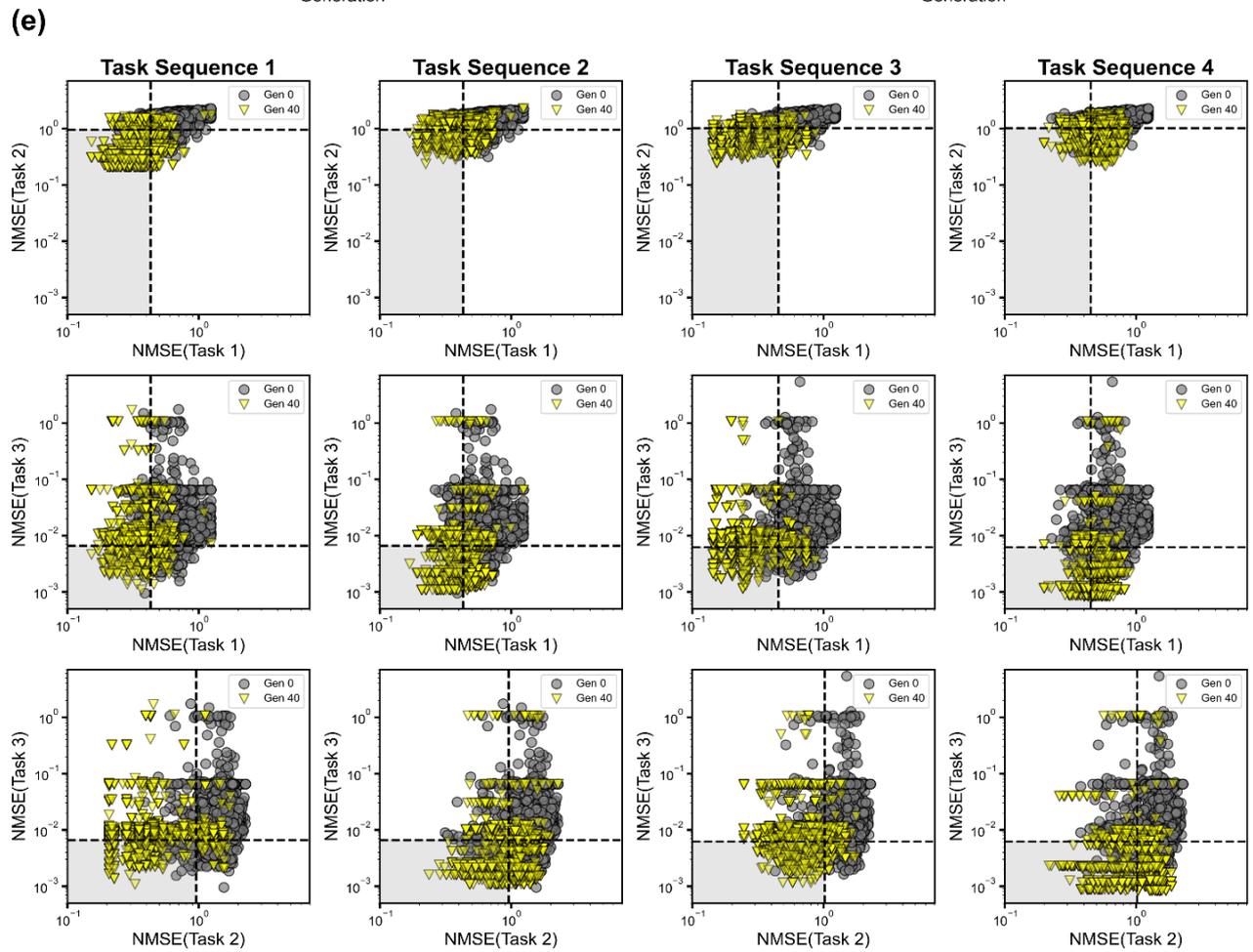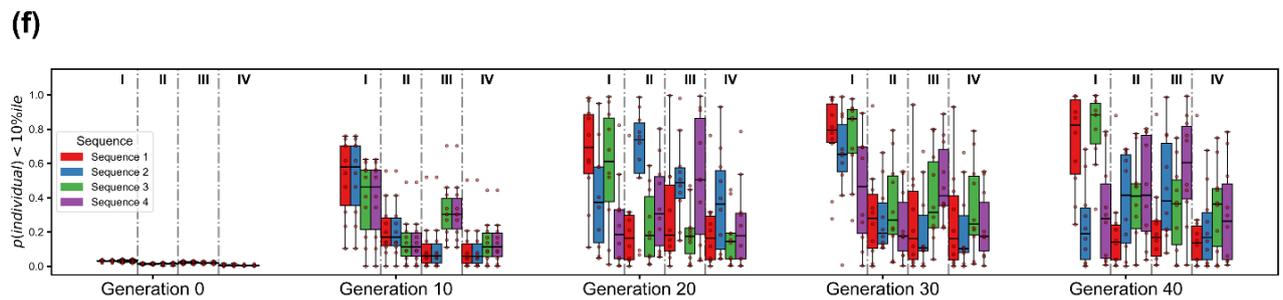

**Figure 3.** Populations of 300 randomly initialized networks were evolved for 10 generations per task segment. **(a)-(d)** The average NMSE of the population across generations for four different tasks series as indicated in the panels. **(e)** Task-correlation plots at Generation 0 (gray circles), and Generation 40 (yellow inverted triangles). **(f)** Probability for individual reservoir networks performing above the 90th percentile on multiple tasks: Task 1 and Task 2 (I); Task 1 and Task 3 (II); Task 2 and Task 3 (III); Task 1, Task 2 and Task 3 (IV).

**Conclusions**

We have studied simulations of a DNA-bead network for its ability to serve as a physical substrate for reservoir computing. We found that even small networks provide sufficient complexity for three distinct tasks, each chosen to impose different requirements on network structure and dynamics. Task performance strongly depends on network topology, and random search was ineffective in identifying networks performing well across all three tasks. Evolution of network structure based on task-based performance was found to be effective at selecting well-performing networks with sparse connectivity. This aligns with recent results by Yadav et al., who considered evolution and performance based selection of reservoir networks but used more generic reservoir dynamics without a direct corresponding physical reservoir structure.(16) Our focus was on reservoir structures transferable into a physical system where DNA acts both as a nonlinear spring and as an information-storage element encoding network structure. We considered a relatively small population of 300 individuals and evolution over 10–40 generations, conditions realistic for implementation using standard genetic tools for *in-vitro* directed evolution. We demonstrated that directed evolution with sequences of heterogeneous tasks was more effective than random search in selecting well-performing networks for varying tasks. Our research highlights how DNA can serve as a memory element useful for computation without the need for explicit "read-out", unlike in other DNA data-storage systems. More broadly, our work bridges natural evolution and its technological counterpart—directed evolution—in the context of learning and computing using physical substrates. Although these results are encouraging, experimental implementation will need to address multiple factors not considered here, such as stability of the network, in particular non-planar structures, efficiency of the linker generation and tracking of the beads. We are currently working on these aspects.

**Methods**

*DNA-bead spring reservoir*

Our simulation and reservoir computing framework are based on the work of Hauser et al. with modifications. Unlike to the work of Hauser et al the springs were described by the wormlike chain (WLC) model, which approximates the force-extension behavior of DNA strands with about 15% relative error.(38)

$$(3)\ F(z_{ij}) = \frac{k_B T}{l_p}\left[\frac{1}{4\left(1-\frac{z_{ij}}{l_c}\right)^2} - \frac{1}{4} + \frac{z_{ij}}{l_c}\right]$$

Where $z_{ij}$ is the extension of the spring between two connected beads $b_i$ and $b_j$. For each studied system the DNA strand was initially at its rest position by placing of the bead and setting $z_{ij} = 0$. The contour length $l_c$ was drawn from a range of 1 μm (approximately 1500 bases) and 200 μm (approximately 317 $10^3$ bases), the strand single-stranded DNA persistence length was $l_p$ = 4 nm, $T$ = 300 K and $k_B$ the Boltzmann constant.(39) The total feedback force on a bead is the sum of forces from all connected strands and was calculated from $\boldsymbol{F}_{bead} = -\sum F(z_{ij})\,\boldsymbol{n}_{i,j}$, with the unit vector $\boldsymbol{n}_{i,j}$ between beads $b_i$ and $b_j$. The colloidal system was considered at small Reynolds number with Stokes' law acting as drag force. With this the equations of motion for each bead are:

$$(4)\ b\dot{x} = F_x + w_{in}X(t)$$

$$(5)\ b\dot{y} = F_y$$

Where $\dot{x}$ and $\dot{y}$ are the velocities of bead, $F_x$ and $F_y$ were the forces acting on the bead in the corresponding spatial dimensions, $b = 1.67 \times 10^{-7}\,\text{N s m}^{-1}$ was the damping coefficient calculated for a bead of radius $10\,\mu m$ in water at room temperature, and $w_{in}X(t)$ the weighted input. If the bead was the input node, then $w_{in}$ was set to $10\,\text{pN}$, otherwise zero, and if the bead was fixed bead, then $\dot{x}$ and $\dot{y}$ were set to 0. The value of $10\,\text{pN}$ represents a typical force amplitude used for manipulation of DNA-bead networks. The equations were numerically integrated with a time step of $1\,\text{ms}$ using SciPy (version 1.14.1) ode solver.(40)

*Thermal Noise*

In some simulations we introduced thermal noise into the bead dynamics by adding a stochastic force term to each node. The net force on the bead was then $\boldsymbol{F}_{bead} = -\sum F(z_{ij})\,\boldsymbol{n}_{i,j} + \xi_i$ where $\xi_i$ is the Brownian stochastic force. The thermal noise term was sampled from a Gaussian distribution, $\xi_i \sim N(0, \sigma^2), \sigma = \frac{\sqrt{2bk_BT}}{\Delta t}$ with $b$ the damping coefficient, $\Delta t = 1ms$ as the simulation time-step, and $k_BT = 4.11 * 10^{-21}\,J$ corresponding to room temperature.

*Reservoir Evaluation and Training*

This interconnected network of DNA-beads springs reservoir was used a mechanical reservoir, where an input force $X(t)$, was applied on the input bead. The system's response was quantified by tracking the time-varying extension $z_{ij}$ of the DNA springs in the reservoir, with the dynamic state matrix (reservoir output matrix) $M(t) \in \mathbb{R}^{T \times N}$ where $T$ is the number of time steps and $N$ is the number of springs. The reservoir output matrix was scalar multiplied by the weight matrix($w = [w_1\ w_2\ w_3\ \ldots\ w_N]^T$) yielding the reservoir output $Y(t) = M(t) \cdot w$. During training, the weights were initially set to unity and subsequently optimized using linear regression against the target signal $\hat{Y}$ (scikit-learn version 1.5.2).(41) The training data for all tasks consisted of a total of 250 000 timesteps and for tasks 1 and 2 the first 80 000 steps were discarded to improve the regression convergence. Post training, the performance was quantified using the normalized mean squared error NMSE = $\frac{1}{N}\sum\frac{(Y_i-\hat{Y}_i)^2}{\bar{Y}\bar{\hat{Y}}}$. Between reservoir output $Y(t)$ and the target signal $\hat{Y}(t)$, where $\bar{Y}$ is the mean of the reservoir output and $\bar{\hat{Y}}$ is the mean of the target signal.

*Genetic Algorithm implementation*

We implemented the evolutionary search using the DEAP (version 1.4.1) library.(42) The initial population of coding strands was generated by randomly selecting nucleotides with equal probability up to the specified sequence length. The genetic algorithm consisted of tournament selection, one-point crossover, and point mutations on DNA sequences to evolve the population over 40 generations in total. The genetic algorithm was implemented with a crossover probability (cxpb), mutation probability for an individual (mutpb), and an independent mutation probability for each allele within an individual. Evaluation was parallelized across CPU cores, and invalid individuals (unstable or disconnected networks) were penalized with high NMSE values of 10000. The parameters used for the genetic algorithm are given in Table 6 in the Supplementary Information unless specified differently.

*Graph analysis*

Graph connectivity and entropy were calculated following earlier works(17) using the python NetworkX package (v3.5).

**Code and Data Availability Statement**

The simulation code can be found online at https://github.com/Bio-inspired-Computation-Lab/evodirect_reservoir. Generated data can be accessed via doi:10.5281/zenodo.17046628.

## Author Contributions

TP (Data analysis, investigation, computer code generation, writing & review of the manuscript), PF (data analysis, writing & review of the manuscript), JS (Conceptualization, methodology, data analysis, investigation, writing & review of the manuscript)

## Acknowledgments

JS would like to acknowledge fruitful discussions with Wilhelm Braun. Funded by the European Union (ERC, SYNNEURO, 101163768). Views and opinions expressed are however those of the author(s) only and do not necessarily reflect those of the European Union or the European Research Council. Neither the European Union nor the granting authority can be held responsible for them. Generative AI tools were used for proofreading of the text.

**Supporting Information**

Supporting Information file with Figures S1 – S7 and SI tables 1 to 6.

Supplementary Material

Directed evolution effectively selects for DNA based physical reservoir computing networks capable of multiple tasks

Tanmay Pandey[1,2], Petro Feketa[3,4], Jan Steinkühler[1,4*]

[1] Bio-Inspired Computation, Institute of Electrical and Information Engineering, Kiel University, Kiel 24143, Germany

[2] Department of Biological Sciences, Indian Institute of Science Education and Research, Mohali, Knowledge City, SAS Nagar, Manauli PO 140306, India

[3] Chair of Automation and Control, Institute of Electrical and Information Engineering, Kiel University, Kiel 24143, Germany

[4] Kiel Nano, Surface and Interface Science KiNSIS, Kiel University, Kiel, Germany

* Corresponding author jst@tf.uni-kiel.de

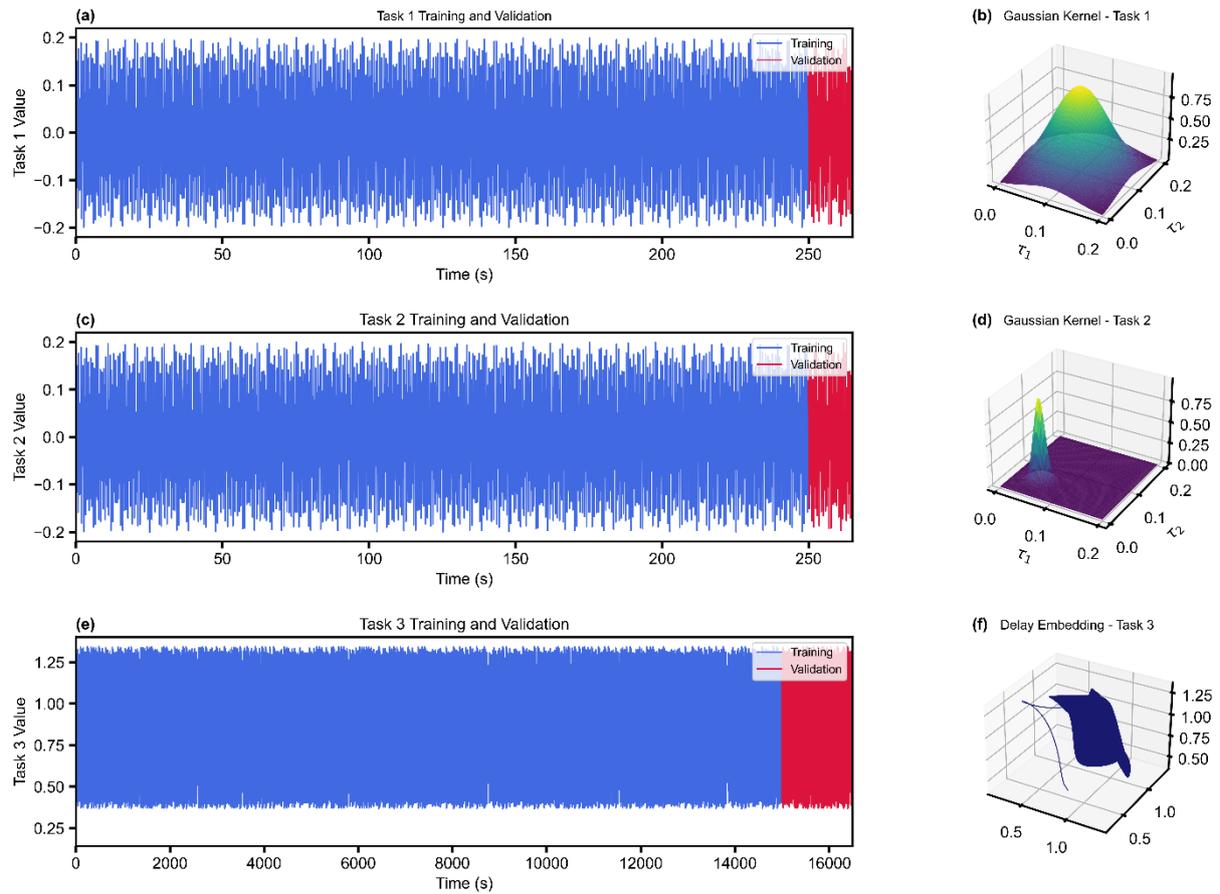

**SI Figure 1**. The sample trajectories of the three tasks: (a) Task 1, (c) Task 2, and (e) Task 3, where the signal used for the training part is in blue, and the validation part is in red. The Gaussian kernel for Task 1 is shown in (b), and that for Task 2 is shown in (d). The delay embedding plot for Task 3 is shown in (f).

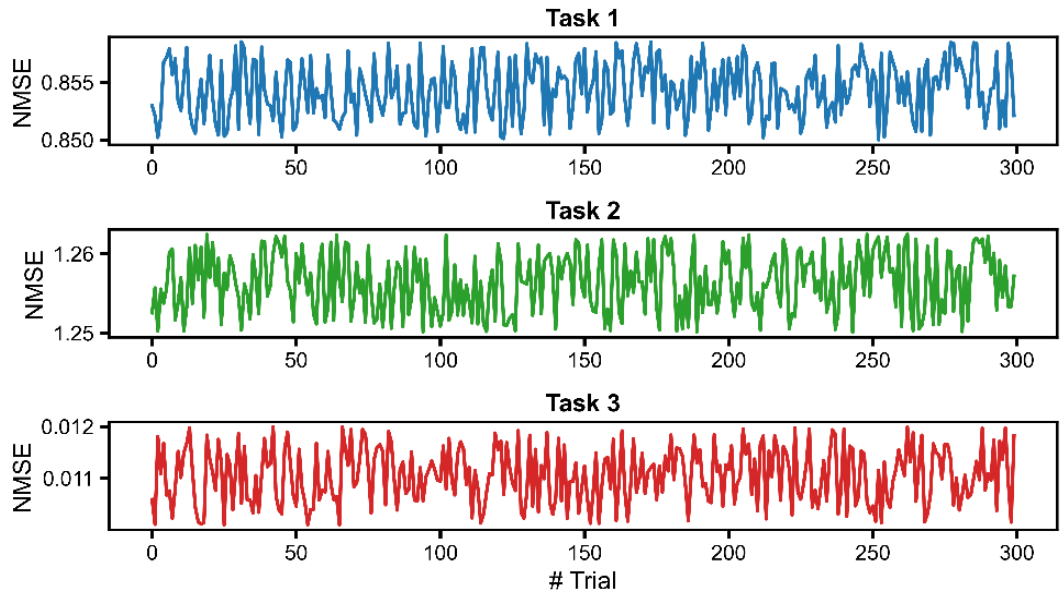

**SI Figure 2**. A system with fixed topology was tested for multiple trials for all the three tasks with different spring contour length combination combinations for each trial, and the predicted NMSE is reported. The system NMSE does not depend strongly on the exact spring values.

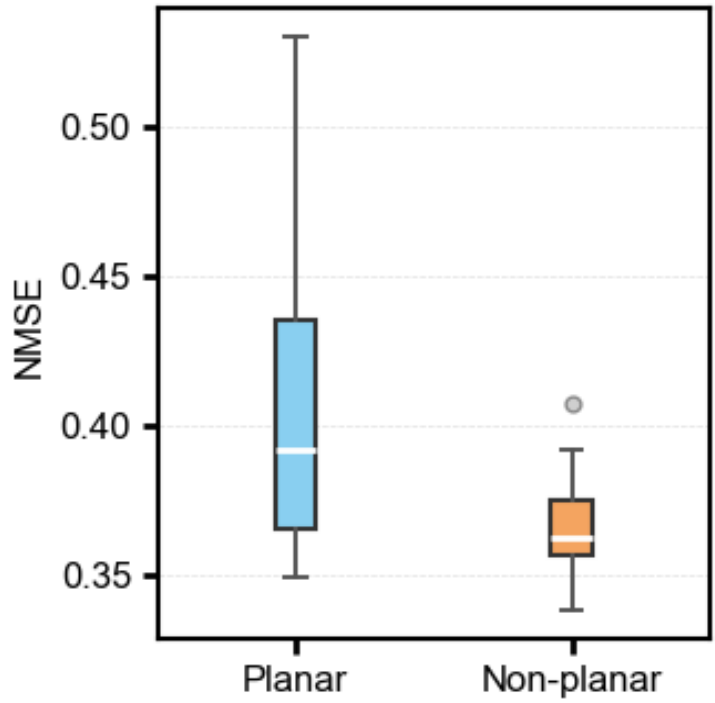

**SI Figure 3.** Box plot of NMSE for ten different planar and non-planar networks that were evaluated on Task 1. The planar network connectivity was coded by Delaunay triangulation. For non-planar networks, all the nodes were connected to each other.

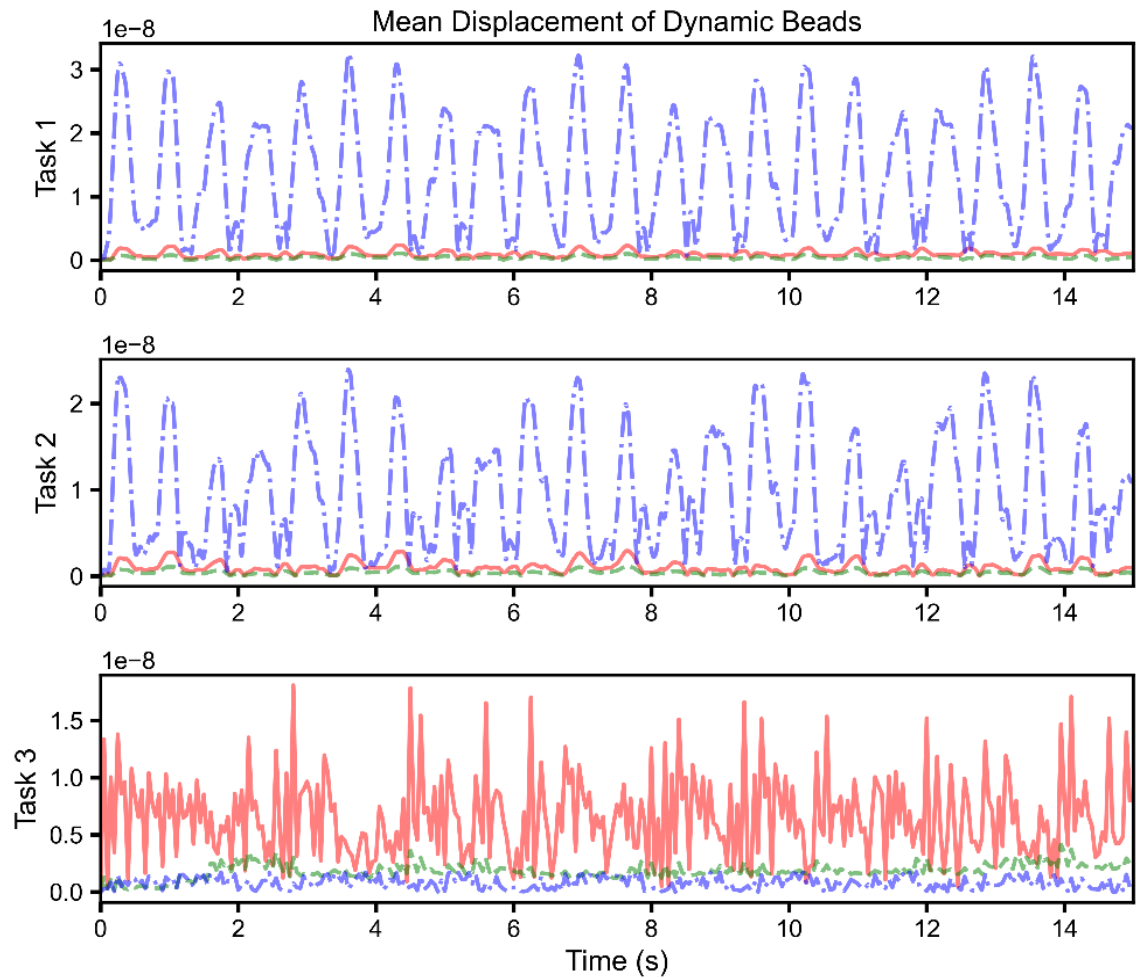

**SI Figure 4.** The bead-displacement for Task 1, Task 2, and Task 3. The red solid line demonstrates the displacement for the average (median NMSE) performing individual, the green dashed line is for the worst (highest NMSE) performing system, and the blue dashed line is for the best (lowest NMSE) performing individual on that task.

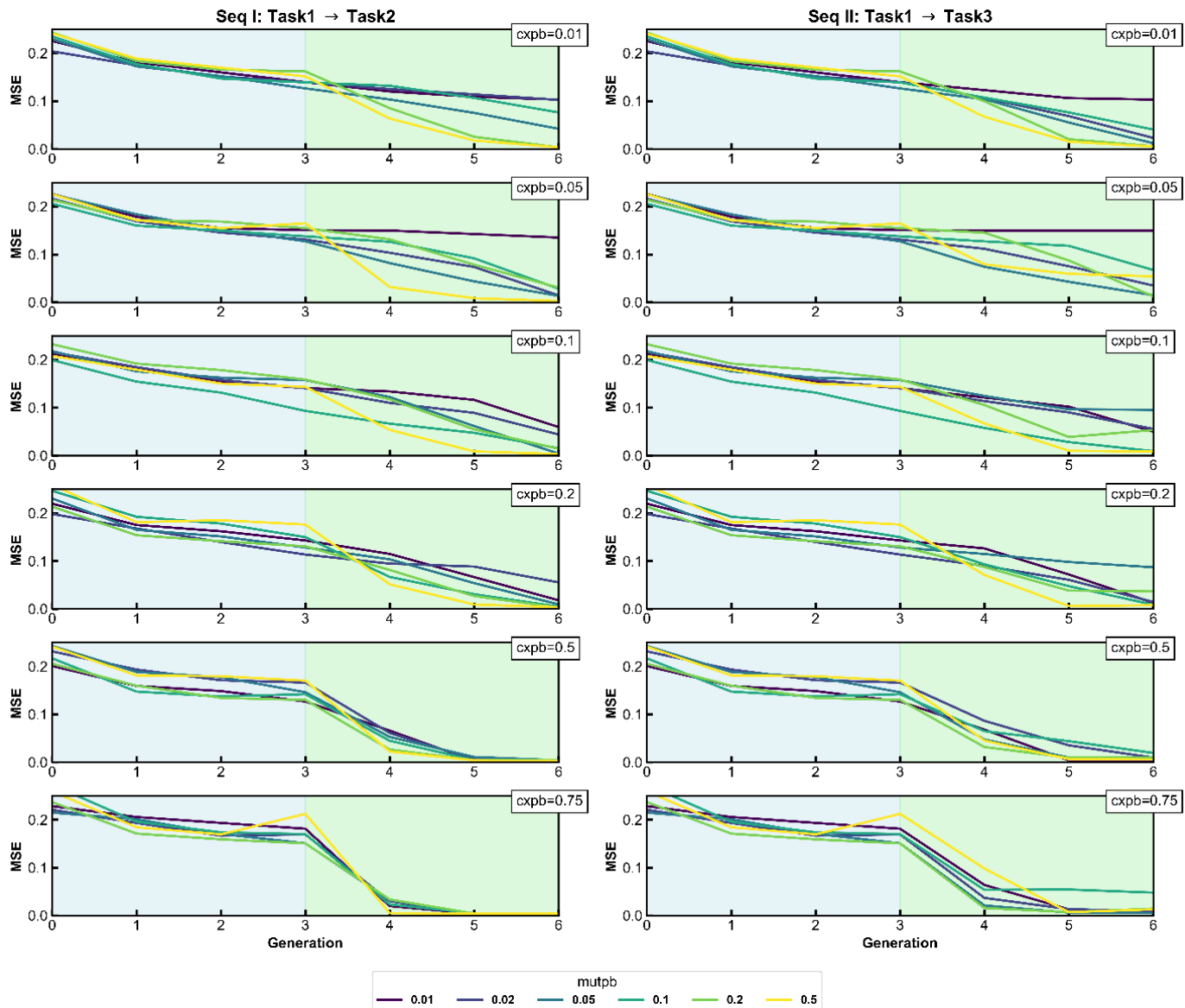

**SI Figure 5a**. A random population of size 30 was initialized and evolved on two different task sequences: Sequence I (Task 1 → Task 2) and sequence II (Task 1 → Task 3), across various combinations of mutation probabilities (mutpb, shown in legend) and crossover probabilities (cxpb, shown as subplot labels). Each subplot shows the mean squared error (MSE) across generations. Shaded regions denote different tasks within each sequence. Parameter combinations leading to smoother (more linear) transitions between tasks are preferred and used in subsequent analyses.

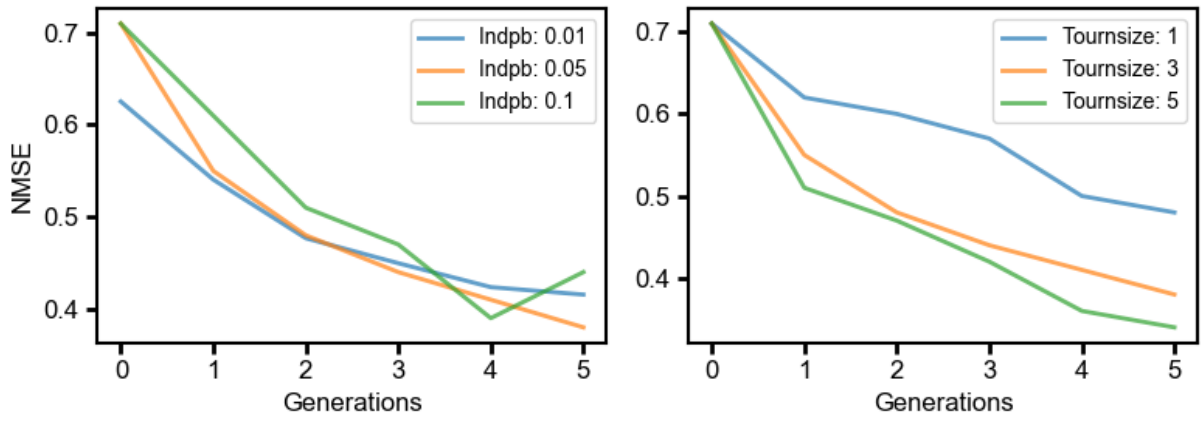

**SI Figure 5b.** A comparison for Genetic Algorithm for two parameters: (i) Variation of nucleotide mutation probability (indpb) and (ii) variation of tournament size (tournsize) in DEAP.

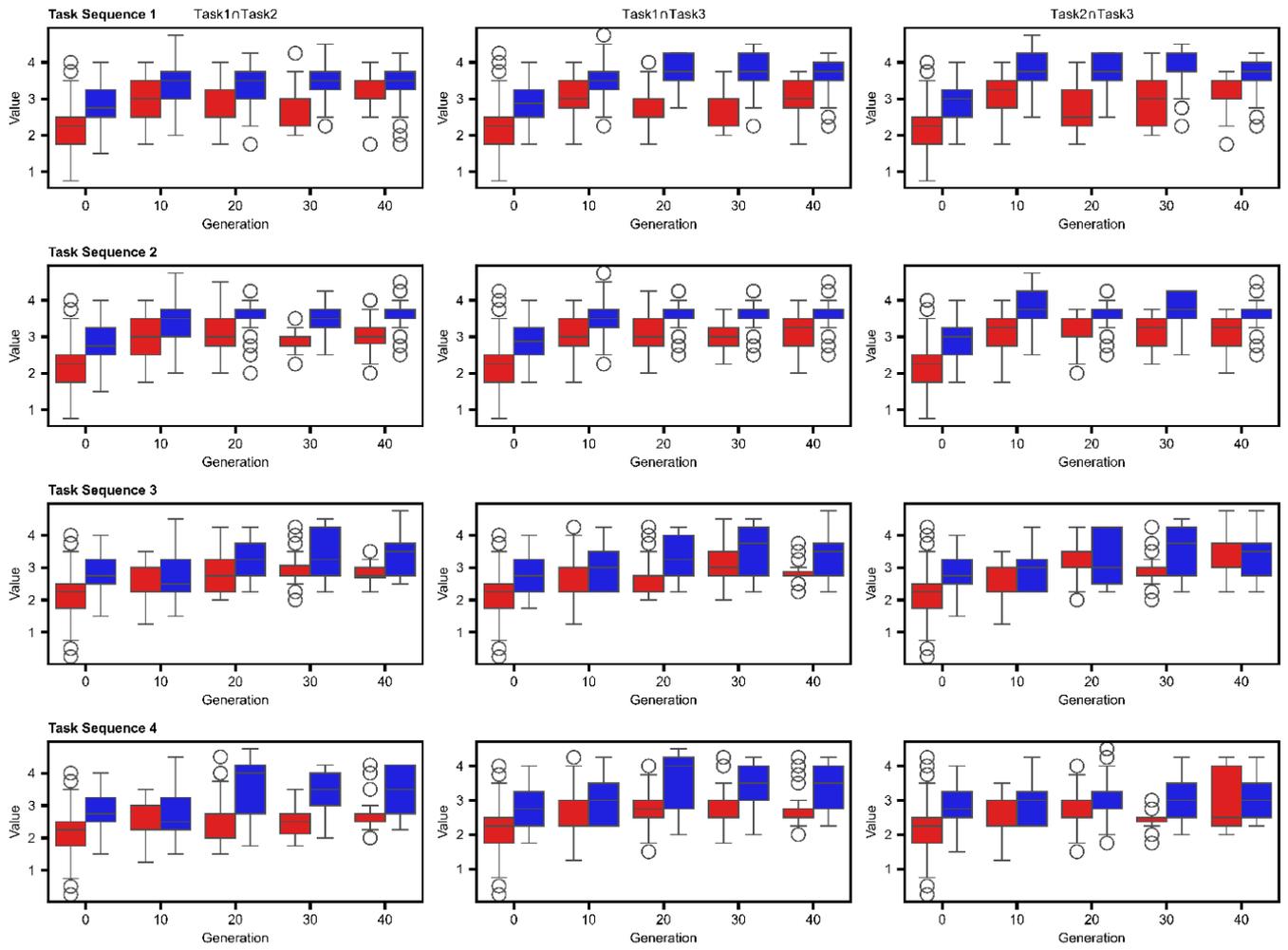

**SI Figure 6.** Network average connectivity analysis of the best-performing individuals from task-correlation quadrant 3 (q3), and the worst performing from task-correlation plot quadrant 1 (q1).

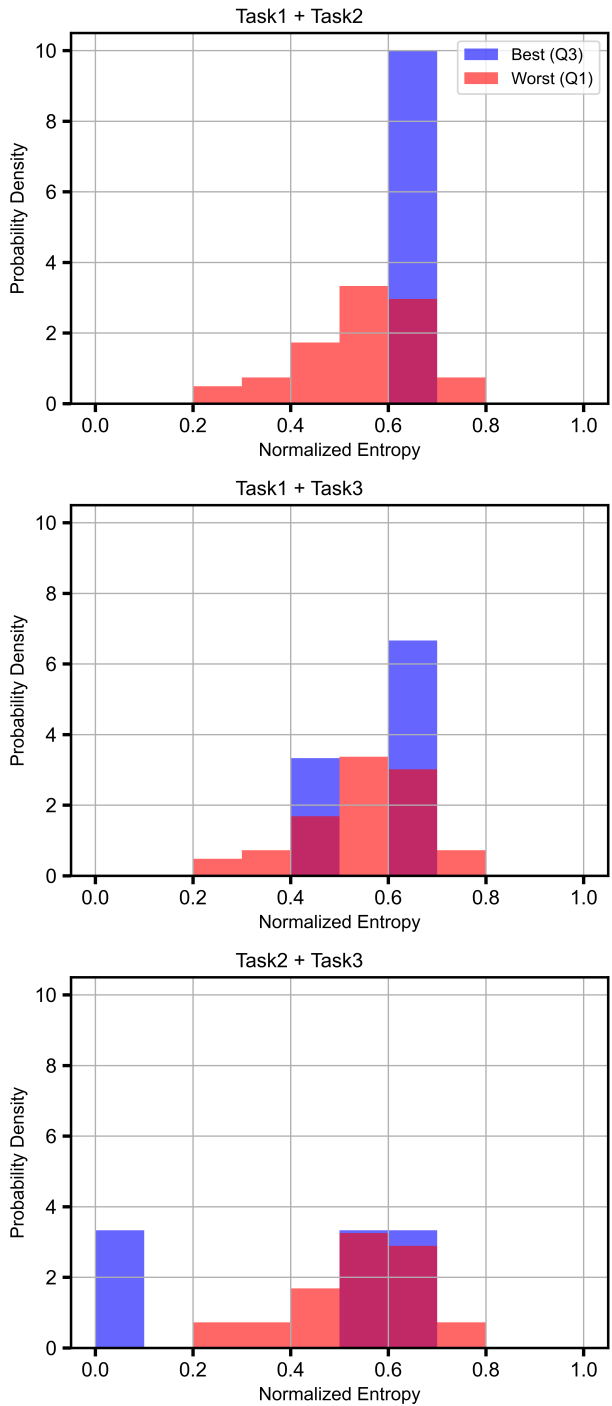

**SI Figure 7.** Normalized network entropy for varying task selection evaluation. Title of each panel indicates that task-correlation quadrant 3 (q3), and worst performing from task-correlation plot quadrant 1 (q1) were deduced from Task A + B.

|  | Liu & Parhi (2022) | Wang & Cichos (2024) | Yadav, Sinha & Stender (2025) | Paul Ahavi et al (2025) | Cherry & Qian (2025) | **Pandey, Feketa & Steinkühler (This work)** |
|---|---|---|---|---|---|---|
| **Substrate** | DNA strand computing | Active polymer beads | Dynamic nodes | Bacterial population | DNA strand computing | Bead-DNA networks |
| **Architecture** | Reservoir computing | Reservoir computing | Reservoir computing | Reservoir computing | Neural network | Reservoir computing |
| **In-material implementation** | yes | yes | no | yes | yes | yes |
| **Evolvable in-material** | no | no | no | yes | no | yes |
| **Learning in-material** | no | no | no | no | yes | no |
| **Network design** | Top down (CAD) | Top down | Bottom up | Bottom up | Top down (CAD) | Bottom up |
| **System size** | 14,000–28,000 DNA reactions | 2 nodes | 10–500 nodes | Bulk system | 1,200 DNA strands | 12 DNA strands* |
| **Timescale for physical interference** | Hours | Real-time** | - | Hours | Hours | Real-time** |
| **Tasks studied** | Classification | Time series | Time series | Classification | Classification | Time series |

**SI Table 1**. Comparison of this work to related computing systems. *Assuming 8 nodes and an average of 4 connections. **Real-time operation is limited by the time constant of the physical system (e.g. due to viscosity) but each cited system operates on much faster timescales than hours.

|  | $\mu_1$ | $\mu_2$ | $\sigma_1$ | $\sigma_2$ | $\Delta t$ |
|---|---|---|---|---|---|
| **Task 1** | 0.1 | 0.1 | 0.05 | 0.05 | 0.001 |
| **Task 2** | 0.05 | 0.05 | 0.01 | 0.01 | 0.001 |

**SI Table 2.** The parameters used to generate the Volterra dataset for task 1 and task 2 with kernel $h_2(\tau_1, \tau_2) = \exp\left(\frac{(\tau_1 - \mu_1)^2}{2\sigma_1^2} + \frac{(\tau_2 - \mu_2)^2}{2\sigma_2^2}\right)$. $U(t) = \sin(2\pi f_1 t) \cdot \sin(2\pi f_2 t) \cdot \sin(2\pi f_3 t)$ with $f_1 = 2.11$, $f_2 = 3.73$ and $f_3 = 4.33$ Hz, $\hat{Y}(t)$ was the target signal and scaling parameter $A = 10^{-11}$ was fixed.

|  | $a$ | $b$ | $c$ | $\tau$ | $\Delta t$ |
|---|---|---|---|---|---|
| **Task 3** | 0.2 | 0.1 | 10 | 17 | 0.1 |

**SI Table 3.** The parameter used to generate the Mackey Glass dataset for task 3.

|  | Task 1 | | Task 2 | | Task 3 | |
| --- | --- | --- | --- | --- | --- | --- |
|  | Brownian Noise | No Brownian Noise | Brownian Noise | No Brownian Noise | Brownian Noise | No Brownian Noise |
| NMSE | 1.0511 | 1.1539 | 0.9434 | 0.9028 | 0.0853 | 0.0700 |

**SI Table 4.** The comparison table of a system with and without Brownian motion tested for all three tasks. Brownian noise was added as Gaussian force noise with variance $2bk_BT/\Delta t$, where $b$ is the drag coefficient at room temperature.

|  | Number of systems | Percentage |
| --- | --- | --- |
| Task 1 ∩ Task 2 | 25 | 2.50% |
| Task 2 ∩ Task 3 | 16 | 1.60% |
| Task 3 ∩ Task 1 | 16 | 1.60% |
| Task 1 ∩ Task 2 ∩ Task 3 | 3 | 0.30% |

**SI Table 5.** Number of systems in the intersection of top 10 % of best performing networks, i.e., having lowest MSE of two or three different tasks.

| Cross-over probability (cxpb) | Mutation probability (mutpb) | Population size (popsize) | Sequence size | Individual probability (indpb) | Tournament Size (tournsize) |
| --- | --- | --- | --- | --- | --- |
| 0.1 (10%) | 0.02 (2%) | 300 | 300 | 0.05 | 3 |

**SI Table 6.** DEAP genetic algorithm parameters